\documentclass{elsarticle}

\expandafter\let\csname equation*\endcsname\relax
\expandafter\let\csname endequation*\endcsname\relax

\usepackage{amsmath}
\usepackage{amssymb}
\usepackage{epstopdf}
\usepackage[usenames, dvipsnames]{color}
\usepackage{graphicx}
\usepackage{graphics}
\usepackage{amsfonts}
\usepackage{eufrak}
\usepackage{mathrsfs}
\usepackage{bm}
\newcommand{\scs}{\scriptscriptstyle}

\begin{document}

\title{Non-Markovian master equation for quantum transport of fermionic carriers}
\author[SFU,IF]{Dmitrii N. Maksimov\corref{FN1}\fnref{FN2}}
\cortext[FN1]{Corresponding author}
\fntext[FN2]{Tel.: +7 391 2494538, Fax: +7 391 2438923}
\ead{mdn@tnp.krasn.ru}
\author[IF,SFU2]{Andrey R. Kolovsky}
\address[SFU]{IRC SQC, Siberian Federal University, 660041, Krasnoyarsk, Russia}
\address[IF]{Kirensky Institute of Physics, Federal Research Center KSC SB
RAS, 660036, Krasnoyarsk, Russia}
\address[SFU2]{ School of Engineering Physics and Radio Electronics,
Siberian Federal University, 660041, Krasnoyarsk, Russia}
\date{\today}
\begin{abstract}
We propose a simple, yet feasible, model for quantum transport of fermionic carriers across
tight-binding chain connecting two reservoirs maintained at arbitrary temperatures and chemical potentials. The model
allows for elementary derivation of the master equation for the reduced single particle density
matrix in a closed form in both Markov and Born approximations. In the Markov approximation the master equation
is solved analytically, whereas in the Born approximation the problem is reduced to an algebraic equation for the single particle
density matric in the Redfield form.
The non-Markovian equation is shown to lead to resonant transport similar to Landauer's conductance.
\end{abstract}
\maketitle

\section{Introduction}

Recently, we have witnessed a lot of interest to quantum transport  \cite{Dubi09a, Znidaric10,Bruderer12,ivanov2013bosonic,
Nietner14,prosen2014exact,Simpson14,Kordas15,Znidaric15, Buca17, Xhani20} across systems
connecting two atomic reservoirs (batteries) \cite{Zozulya13, Caliga17}. Specifically, such system can nowadays be set
experimentally with ultracold atomic gases \cite{Bran12,labouvie2015negative,lebrat2018band}. One of the major tools for theoretical
analysis of such systems is the master equation approach \cite{Pershin08, Pepino10, Dale14, landi2021non}. Despite the enormous progress,
so far the
approach has been fully established only in the framework of the Born-Markov approximation
\cite{Pepino10,Dale14,Nietner14,prosen2014exact,Xu17,kolovsky2018landauer}.
To handle the non-Markovian
regimes for fermionic carriers the stochastic Schr\"odinger equation approaches with the correlated noise
\cite{Diosi98, Zhao12, Chen13, Suess15} has been put forward. As it is demonstrated in \cite{Suess15},
these approaches result in a hierarchy of stochastic evolution equations of the diffusion type with
Grassmannian noise making
it difficult to simulate numerically due to anticommutative multiplication. In this paper
we analyse a model for quantum transport of fermionic carriers recently proposed in \cite{kolovsky2021resonant}.
We will show that the model
allows for elementary derivation of a numerically tractable non-Markovian master equation in a closed form whereas in
the Markov approximation
the model is solvable analytically.

\begin{figure}[t]
\begin{center}
{\includegraphics[width=0.5\textwidth,height=0.15\textwidth,trim={4.8cm
13.5cm 4.0cm 12.5cm},clip]{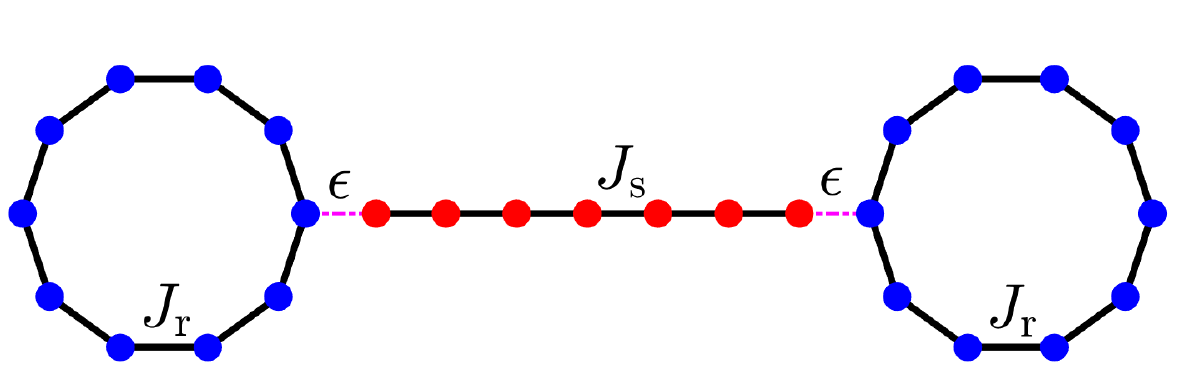}} \caption{Sketch of the set-up.} \label{fig1}
\end{center}
\end{figure}

We consider the set-up consisting of a linear tight-binding chain of $L$ sites coupled at
both ends with two tight-binding rings of $M$ sites each \cite{Kolovsky20a,kolovsky2021resonant} as shown in Fig. \ref{fig1}.
Throughout the text the chain is termed {\it system}, whereas the rings
are going to be referred to as {\it reservoirs}. Non-interacting spinless fermions can move between
the sites of the system and the sites of the reservoirs with hopping rates $J_{\rm s, r}$, correspondingly.
The hopping between the system and the reservoirs is quantified by the coupling constant $\epsilon$.
The dynamics is controlled
by the master equation for the total density matrix
\begin{equation}\label{Master_full}
\frac{\partial \widehat{{\cal R}}}{\partial t}=-i[\widehat{{\cal H}},  \widehat{{\cal R}}]+
\gamma\sum_{\ell=1,L}\sum_{\nu=1}^{M}\left(\widehat{{\cal L}}^{\scs (g)}_{\ell,\nu}+\widehat{{\cal L}}^{\scs (d)}_{\ell, \nu}\right).
\end{equation}
The Hamiltonian in Eq. (\ref{Master_full}) is written as
\begin{align}\label{Ham_tot}
\widehat{{\cal H}}=\widehat{{\cal H}}_{\rm s}+\sum_{\ell=1,L}\left(
\widehat{{\cal H}}_{{\rm r},\ell}+\widehat{{\cal H}}_{{\rm c},\ell}\right),
\end{align}
where
\begin{align}\label{Ham_sys}
\widehat{{\cal H}}_{\rm s}=
-\frac{J_{\rm s}}{2}\sum_{\ell=1}^{L-1}\hat{a}_{\ell+1}^{\dagger}\hat{a}_{\ell} +{\rm h.c.}
\end{align}
is the system's Hamiltonian with $\hat{a}_{\ell}^{\dagger},\hat{a}_{\ell}$
being Fermionic
creation and annihilation operators at the $\ell_{\rm th}$ site.
The reservoirs'  $\widehat{{\cal H}}_{{\rm r},\ell}$ and the coupling  $\widehat{{\cal H}}_{{\rm c},\ell}$ Hamiltonians
are indexed with
subscript $\ell$ specifying
the connection site. For further convenience we write each reservoir Hamiltonian in terms of Fermionic
operators acting in the Fock space of the Bloch eigenstates of the ring
\begin{align}\label{Ham_res}
& \widehat{{\cal H}}_{\rm r}=
-J_{\rm r}\sum_{\nu=1}^{M}\cos(k_{\nu})\hat{a}_{\nu}^{\dagger}\hat{a}_{\nu}, \nonumber \\
& k_{\nu}=\frac{2\pi\nu}{M}, \ \nu=1,2,\ldots, M,
\end{align}
where $\nu$ enumerates the Bloch eigenstates. We dropped subscript $\ell$ in Eq.~(\ref{Ham_res})
assuming that the reservoirs are identical. The coupling Hamiltonian
is given by
\begin{align}\label{Ham_cou}
\widehat{{\cal H}}_{\rm c, \ell}=
-\frac{\epsilon}{2\sqrt{M}}\sum_{\nu=1}^M\hat{a}_{\ell}^{\dagger}\hat{a}_{\nu} +{\rm h.c.}.
\end{align}

To prescribe thermodynamic quantities to each
reservoir we introduced the particle drain
\begin{equation}\label{drain}
\widehat{{\cal L}}^{\scs(d)}_{\ell, \nu}=\frac{\bar{n}_{\nu,\ell}-1}{2}
\left(\hat{a}_{\nu}^{\dagger}\hat{a}_{\nu}\widehat{\cal R }-2\hat{a}_{\nu}\widehat{\cal R }\hat{a}_{\nu}^{\dagger}
+\widehat{\cal R }\hat{a}_{\nu}^{\dagger}\hat{a}_{\nu} \right),
\end{equation}
and the particle gain
\begin{equation}\label{gain}
\widehat{{\cal L}}^{\scs{(g)}}_{\ell, \mu}=-\frac{\bar{n}_{\nu,\ell}}{2}
\left(\hat{a}_{\nu}\hat{a}_{\nu}^{\dagger}\widehat{\cal R }-2\hat{a}_{\nu}^{\dagger}\widehat{\cal R }\hat{a}_{\nu}
+\widehat{\cal R }\hat{a}_{\nu}\hat{a}_{\nu}^{\dagger} \right)
\end{equation}
Lindblad operators \cite{landi2021non}, where
\begin{equation}
\bar{n}_{\nu,\ell}= \frac{1}{e^{-\beta_{\ell}[J_{\rm r}\cos(k_{\nu})+\mu_{\ell}]}+1}
\end{equation}
ensures that due to coupling with Lindbald bath \cite{Dubi09, Ajisaka12, nazir2018reaction} each reservoir is populated according to the Fermi-Dirac distribution
$n=(e^{\beta(E-\mu)}+1)^{-1}$
with given chemical potential $\mu_{\ell}$ and inverse temperature
$\beta_{\ell}$. The reservoirs are labelled by their connection site $\ell=1,L$. Finally, the constant $\gamma$ in Eq.~(\ref{Master_full}) is the reservoir relaxation rate
which determines how fast the isolated reservoir relaxes to the thermodynamic equilibrium.

\section{Single particle density matrix}
Equation (\ref{Master_full}) only contains pairwise combinations of the creation and annihilation operators.
This allows us to rewrite it in terms of the total single particle density matrix (SPDM) $\hat{\rho}$.
The entries of the SPDM are defined as follows
\begin{equation}\label{SPDM}
\rho_{q',q}=\mathrm{tr} (\hat{a}_{q}^{\dagger}\hat{a}_{q'}\widehat{{\cal R}}),
\end{equation}
where $q$ spans all Bloch degrees of freedom $\nu$ in the reservoirs as well as the Wannier degrees of
freedom $\ell$ in the system. Let us assume for a moment that only one reservoir is attached to the system
at $\ell=1$, then the SPDM takes the following block form
\begin{equation}\label{SPDM_block}
\hat{\rho}=
\left(
\begin{array}{cc}
\hat{\rho}_{\rm r} & \hat{\rho}_{\rm c} \\
\hat{\rho}_{\rm c}^{\dagger} & \hat{\rho}_{\rm s}
\end{array}
\right),
\end{equation}
where $\hat{\rho}_{\rm r}$ is the SPDM of the reservoir, $\hat{\rho}_{\rm s}$ is the SPDM of the system,
and  $\hat{\rho}_{\rm c}$ accounts for reservoir-system correlations.
The following three equations can be obtained by applying Eq.~(\ref{SPDM_block}) to Eq.~(\ref{Master_full})
\begin{align}
& \frac{\partial \hat{\rho}_{\rm s}}{\partial t}=-i[\widehat{H}_{\rm s}, \hat{\rho}_{\rm s}]
-i\epsilon(\widehat{V}^{\dagger}_1\hat{\rho}_{\rm c}-\hat{\rho}_{\rm c}^{\dagger}\widehat{V}_1), \label{one} \\
& \frac{\partial \hat{\rho}_{\rm c}}{\partial t}=-i\widehat{H}_{\rm r}\hat{\rho}_{\rm c}
+i\hat{\rho}_{\rm c}\widehat{H}_{\rm s}
-\frac{\gamma}{2}\hat{\rho}_{\rm c}
-i\epsilon(\widehat{V}_1\hat{\rho}_{\rm s}-\hat{\rho}_{\rm r}\widehat{V}_1), \label{two} \\
& \frac{\partial \hat{\rho}_{\rm r}}{\partial t}=-i[\widehat{H}_{\rm r}
,\hat{\rho}_{\rm r}]
-i\epsilon(\widehat{V}_1\hat{\rho}_{\rm c}^{\dagger} -\hat{\rho}_{\rm c}\widehat{V}^{\dagger}_1)
+\gamma(\hat{\rho}^{\scs (0)}_{ \rm r}-\hat{\rho}_{\rm r}), \label{three}
\end{align}

where $\hat{\rho}^{\scs (0)}_{ \rm r}$ is the Fermi-Dirac SPDM of the reservoir
\begin{equation}
\hat{\rho}^{\scs (0)}_{ \rm r}=\sum_{\nu=1}^M\frac{|\nu\rangle\langle\nu|}{e^{-\beta[J_{\rm r}\cos(k_{\nu})+\mu]}+1},
\end{equation}
while the Hamiltonian
\begin{equation}\label{Ham_single}
\widehat{H}=
\left(
\begin{array}{cc}
\widehat{H}_{\rm{r}} & \epsilon\widehat{V}_1 \\
\epsilon\widehat{V}^{\dagger}_1 & \widehat{H}_{\rm{s}}
\end{array}
\right)
\end{equation}
is composed of the single particle Hamiltonian of the system
\begin{equation}\label{Ham_single_sys}
\widehat{H}_{\rm{s}}=-\frac{J_{\rm s}}{2}\sum_{\ell=1}^{L-1}\left( |1\!+\!\ell\rangle\langle \ell|+\rm {h.c.}\right),
\end{equation}
the single particle Hamiltonian of the reservoir
\begin{equation}\label{Ham_single_res}
\widehat{H}_{\rm{r}}=-J_{\rm{r}}\sum_{\nu=1}^M \cos\left(\frac{2\pi \nu}{M}\right)|\nu\rangle \langle \nu|,
\end{equation}
and the coupling operator
\begin{equation}\label{V}
\widehat{V}_{\ell}=-\frac{1}{2\sqrt{M}}\sum_{\nu=1}^M|\nu\rangle\langle\ell|.
\end{equation}
From Eq.~(\ref{two}) we find the solution with the initial condition $ \hat{\rho}_{\rm c}(0)=0$
\begin{equation}\label{c}
\hat{\rho}_{\rm c}\!=\!i\epsilon\!\int\limits_{0}^{t}d\tau
e^{\frac{\gamma}{2}(\tau\!-\!t)}\widehat{U}_{\rm r}(t\!-\!\tau)\!\left[
\hat{\rho}_{\rm r}(\tau)\widehat{V}_1\!-\!\widehat{V}_1\hat{\rho}_{\rm s}(\tau)\right]
\!\widehat{U}_{\rm s}^{\dagger}(t\!-\!\tau),
\end{equation}
where $\widehat{U}_{\rm s,r}(t)=\widehat{\exp}(-i\widehat{H}_{\rm s,r}t)$ are the evolution operators and
the initial condition $\hat{\rho}_{\rm c}(0)=0$ corresponds to
the absence of initial reservoir-system correlations.

The above procedure can be applied to a reservoir attached
to an arbitrary site of the chain. To address the transport problem the
second reservoir is reattached to the $L_{\rm th}$ site.
From now on we apply the notations $\hat{\rho}_{\ell}$ for
the SPDM of the reservoir at the $\ell_{\rm th}$ site.
By substituting Eq.~(\ref{c}) into Eq.~(\ref{one}) and changing variables $\tau-t\rightarrow \tau$
one finds
\begin{equation}\label{master_reduced}
\frac{\partial \hat{\rho}_{\rm s}}{\partial t}=-i[\widehat{H}_{\rm s},\hat{\rho}_{\rm s}]
+\sum_{\ell=1,L}\left(\widehat{L}_{\ell}+
\widehat{L}_{\ell}^{\dagger}\right),
\end{equation}
where

\begin{equation}\label{Ldag}
\widehat{L}_{\ell}\!=\!\epsilon^2\!\int\limits_{-t}^{0}d\tau e^{\frac{\gamma}{2}\!\tau}
\widehat{V}^{\dagger}_{\ell}\widehat{U}_{\rm r}^{\dagger}(\tau)\left[
\hat{\rho}_{\ell}(\tau\!+\!t)\widehat{V}_{\ell}\!-\!\widehat{V}_{\ell}\hat{\rho}_{\rm s}(\tau\!+\!t)\right]\widehat{U}_{\rm s}(\tau).
\end{equation}

\section{Markov approximation}

The Markov approximation consists of assuming no memory in integral Eq. (\ref{c}).
The Markov approximation makes it possible to derive the master equation for $\hat{\rho}_{\rm s}$ as a set
of ordinary differential equations. The elementary derivation
is presented in Appendix A. The final result is

\begin{align}\label{Markovian}
& \frac{\partial \hat{\rho}_{\rm s}}{\partial t}=-i[\widehat{H}_{\rm s}, \hat{\rho}_{\rm s}]
-\frac{\epsilon^2}{2\gamma}\sum_{\ell=1,L}\left\{|\ell \rangle\langle \ell|,\hat{\rho}_{\rm s} \right\}
+\frac{\epsilon^2}{\gamma}\sum_{\ell=1,L}\left(\frac{{\gamma^2\bar{n}_{\ell}}}{\gamma^2+\epsilon^2}+
\frac{{\epsilon^2}}{\gamma^2+\epsilon^2} \langle \ell|\hat{\rho}_{\rm s}|\ell \rangle \right)
|\ell \rangle\langle \ell|,
\end{align}
where
\begin{equation}
\bar{n}_{\ell}=\frac{1}{M}\sum_{\nu=1}^M\frac{1}{e^{-\beta[J_{\rm r}\cos(k_{\nu})+\mu]}+1}
\end{equation}
is the mean population of each site of the reservoir at the $\ell_{\rm th}$ site
in the absence of coupling $\epsilon\!=\!0$
and $\{{\scs \ldots},{\scs \ldots}\}$ designates the anticommutator.
Equation (\ref{Markovian}) can be solved with a three diagonal time-stationary Ansatz
\begin{equation}
\hat{\rho}_{\rm s}\!=\!\sum_{\ell=1,L}A_{\ell}|\ell\rangle\langle\ell|\!+\!B\sum_{\ell=1}^{L-1}
(i|\ell\!+\!1\rangle\langle\ell|+{\rm h.c.}\!)\!+\!
A\sum_{\ell=2}^{L-1}|\ell\rangle\langle\ell|
\end{equation}
which, upon substitution into Eq. (\ref{Markovian}), yields
\begin{align}
& A_{1}=C+\frac{\epsilon^2}{\gamma J_{\rm s}}B, \ A_{L}=C-\frac{\epsilon^2}{\gamma J_{\rm s}}B \nonumber, \\
& B=\frac{1}{2}\cdot\frac{(\bar{n}_{1}-\bar{n}_{L})J_{\rm s}\gamma\epsilon^2}{J_{\rm s}^2(\gamma^2+
\epsilon^2)+\epsilon^4}, \ \ C=\frac{\bar{n}_{1}+\bar{n}_{L}}{2}.
\end{align}
The stationary probability current along any bond in the system can be found
as $\langle j \rangle=JB$. Thus, we have
\begin{equation}\label{current}
\langle j \rangle=\frac{1}{2}\cdot\frac{(\bar{n}_{1}-\bar{n}_{L})J_{\rm s}^2\gamma\tilde{\gamma}}{J_{\rm s}^2
(\gamma+\tilde{\gamma})+\gamma\tilde{\gamma}^2},
\end{equation}
where we introduced
\begin{equation}
\tilde{\gamma}=\frac{\epsilon^2}{\gamma}.
\end{equation}

If $\gamma\gg\epsilon$, Eq. (\ref{Markovian}) simplifies
to
\begin{equation}\label{BM}
\frac{\partial \hat{\rho}_{\rm s}}{\partial t}\!=\!-i[\widehat{H}_{\rm s},
\hat{\rho}_{\rm s}]\!-\tilde{\gamma}\sum_{\ell=1,L}\!
\left(\frac{1}{2}\left\{|\ell \rangle\langle \ell|,\hat{\rho}_{\rm s} \right\}
\!-\!\bar{n}_{\ell}|\ell \rangle\langle \ell|\right)\!.
\end{equation}
The condition $\gamma\gg\epsilon$ implies that the thermalization time of the reservoirs is much shorter than the
time-scale of the dynamics induced by the system-reservoir coupling,
i.e. the system interacts with a quasi-thermalized reservoir.
It is not difficult to see that Eq. (\ref{BM}) is can be derived from the following many particle master equation
for the reduced density matrix $\widehat{{\cal R}}_{\rm s}={\rm tr}_{\rm r}(\widehat{{\cal R}})$
\begin{equation}\label{Bloch_reduced}
\frac{\partial \widehat{{\cal R}}_{\rm s}}{\partial t}=-i[\widehat{{\cal H}}_{\rm s},  \widehat{{\cal R}}_{\rm s}]+
\tilde{\gamma}\sum_{\ell=1,L}\left(\widehat{{\cal L}}_{\ell}^{\scs \rm (g)}+\widehat{{\cal L}}_{\ell}^{\scs \rm (d)}\right),
\end{equation}
with $\tilde{\gamma}$ playing the role of the effective reservoir relaxation rate, and $\widehat{{\cal L}}_{\ell}^{\scs \rm (g,d)}$ the
standard gain and drain Lindblad operators of the form Eq. (\ref{drain}), and Eq. (\ref{gain}), but now acting directly
on the Wannier state of the connection sites. Equation (\ref{Bloch_reduced}) is usually obtained with
application of both Markov and Born approximations \cite{Breu02}. Physically, the Born approximation implies weak coupling between the
system and the reservoir $J_{{\rm s,r}}\ll\epsilon$. Often \cite{Breu02} the Born approximation is introduced as
$ \widehat{{\cal R}}=\widehat{{\cal R}}_{\rm r}
\otimes\widehat{{\cal R}}_{\rm s}$.
It can be easily seen that in the SPDM language the above becomes
${ \hat{\rho}}={\hat{\rho}_{\rm r}}\oplus{\hat{\rho}_{\rm s}}$.
In our case the latter formula does not hold true \cite{Kolovsky20} as it can be easily seen from Eq. (\ref{c}).
In fact, the zeroth order Born approximation emerges as
\begin{equation}\label{Born}
\hat{\rho}_{\ell}=\hat{\rho}^{\scs (0)}_{\ell}
\end{equation}
meaning that the reservoir's SPDM is not perturbed by the state of the system.

\section{Born Approximation}

Let us apply the Born approximation to Eq.~(\ref{Ldag})
not involving the Markov approximation at the initial step.
Substituting Eq.~(\ref{Born}) into Eq.~(\ref{Ldag}) and taking the limit $M\rightarrow\infty$
one finds
\begin{equation}\label{Ldag2}
\widehat{L}_{\ell}\!=\!\frac{\epsilon^2}{4}|\ell\rangle\langle\ell|\!\int\limits_{-t}^{0}\!d\tau
e^{\frac{\gamma}{2}\tau}\!
\left[{\cal J}_{\rm F}(J_{\rm r}\tau)
\widehat{\mathbb I}_{\rm s}\!-\!{\cal J}_{0}(J_{\rm r}\tau)\hat{\rho}_{\rm s}(\tau\!+\!t)
\!\right]\widehat{U}_{\rm s}(\tau),
\end{equation}
where ${\cal J}_0$ is the zeroth order Bessel function of the first kind, $\widehat{\mathbb I}_{\rm s}$ the identity operator in
the Wannier basis of the system, and
\begin{equation}\label{J_Fermi}
{\cal J}_{\rm F}(J_{\rm r} t)=\frac{1}{2\pi}\int\limits_{-\pi}^{\pi}d\kappa\frac{e^{-iJ_{\rm r}\cos(\kappa)t}}
{e^{-\beta[J_{\rm r}\cos(\kappa)+\mu]}+1}.
\end{equation}
Equation (\ref{Ldag2}) together with Eq.~(\ref{master_reduced}) constitute the non-Markovian master equation in
the Born approximation. Notice the key role of $\gamma$ in Eq.~(\ref{Ldag2}); since the Bessel function at large
$t$ decays as $1/\sqrt{t}$ the integral in Eq.~(\ref{Ldag2}) is only convergent
with non-zero $\gamma$.

If the difference between the chemical potentials $\Delta\mu=\mu_1-\mu_L$ is small in comparison to the $\widehat{H}_{\rm s}$ level spacing, the system's SPDM can be written as
\begin{equation}
\hat{\rho}_s=\hat{\rho}_s^{\scs (0)}+\Delta\mu\hat{\rho}_s^{\scs (1)}.
\end{equation}
Note that $\hat{\rho}_s^{\scs (0)}$ corresponds to equilibrium, and, thus, does not
support a probaility current. From Eq.~(\ref{master_reduced}) we have
\begin{equation}\label{master_Land}
\frac{\partial\hat{\rho}_{\rm s}^{\scs (1)}}{\partial t}=-i[\widehat{H}_{\rm s},\hat{\rho}_{\rm s}^{\scs (1)}]
+\sum_{\ell=0,L}\left(\widehat{\Delta}_{\ell}+\widehat{\Delta}_{\ell}^{\dagger}\right).
\end{equation}
At low temperatures, $\beta\gg J_{\rm r}$ the Fermi-Dirac distribution is
\begin{equation}
\lim_{\beta\rightarrow\infty}n(E,\mu+\Delta\mu)=\theta(\mu-E)+{\Delta\mu}\delta(E-\mu),
\end{equation}
where $\theta$ is the Heaviside theta. Thus, for the operators $\widehat{\Delta}_{\ell}$ one finds
\begin{align}\label{delta}
\widehat{\Delta}_{\ell}\!=\!\frac{\epsilon^2}{4}|\ell\rangle\!\langle\ell|\!\int\limits_{-t}^{0}\!d\tau
e^{\frac{\gamma\tau}{2}}\!\left[\!\frac{d(\mu)\delta_{1,\ell}}{M}e^{i\mu\tau}\widehat{\mathbb I}_{\rm s}
\!-\!{\cal J}_{0}(J_{\rm r}\tau)\hat{\rho}_{\rm s}^{\scs (1)}(\!\tau\!+\!t)\!\right]\!\widehat{U}_{\rm s}(\!\tau\!),
\end{align}
where $d(\mu)$ is the $M$-site reservoir density of states
\begin{equation}\label{DOS}
d(\mu)=
\left\{
\begin{array}{cl}
\frac{MJ_{\rm r}}{\pi\sqrt{J_{\rm r}^2-\mu^2}} & \ {\rm if} \ |J_{\rm r}|>|\mu|, \\
0  & \ {\rm if} \ |J_{\rm r}|<|\mu|.
\end{array}
\right.
\end{equation}
%

Finally, let us find the stationary equation for the the matrix $\hat{\rho}_s^{\scs (1)}$.
Using the eigenenergies $E_m$ and eigenstates $|m\rangle$ of $\widehat{H}_{\rm s}$
\begin{align}
&  E_m=-J\cos\left(\frac{\pi m}{L+1}\right), \ m=1, 2, \ldots, L, \nonumber \\
& |m\rangle=\sqrt{\frac{2}{L+1}}\sum_{\ell=1}^L\sin\left(\frac{\pi m\ell}{L+1}\right)|\ell\rangle.
\end{align}
we write both $\hat{\rho}_s^{\scs (1)}$ and $\widehat{\mathbb I}_{\rm s}$ as a series expansion
\begin{align}\label{expansion}
& \hat{\rho}_{\rm s}^{\scs (1)}=\sum_{m,m'=1}^L {\rho}_{m,m'}|m\rangle\langle m'|, \nonumber \\
& \widehat{\mathbb I}_{\rm s}=\sum_{m=1}^L| m\rangle\langle m|.
\end{align}
By substituting Eq.~(\ref{expansion}) into Eq.~(\ref{master_Land})
we obtain
\begin{equation}\label{master_station}
 i\sum_{m,m'=1}^L {\rho}_{m,m'}(E_m-E_{m'})|m\rangle\langle m'|=
\sum_{\ell=1,L}\left[\widehat{\Delta}_{\ell}(t_{\infty})+\widehat{\Delta}_{\ell}^{\dagger}(t_{\infty})\right]
\end{equation}
with
\begin{equation}\label{L_station}
\widehat{\Delta}_{\ell}(t_{\infty})=\frac{\epsilon^2}{4}|\ell\rangle\langle \ell|
\left(\frac{J_{\rm r}\delta_{1,\ell}}{\pi\sqrt{J_{\rm r}^2-\mu^2}}\sum_{m=1}^L \frac{i |m\rangle\langle m|}{E_m-\mu+i\frac{\gamma}{2}}-
 \sum_{m,m'=1}^L \frac{{\rho}_{m,m'}|m\rangle\langle m'|}{\sqrt{J_r^2-\left(E_{m'}+i\frac{\gamma}{2} \right)^2}}\right),
\end{equation}
where we used
\begin{equation}
\int\limits_{-t}^0 d\tau e^{\left(\frac{\gamma}{2}-iE_m\right)\tau}{\cal J}_{0}(J_{\rm r} \tau)=
\frac{1}{\sqrt{J_{\rm r}^2+\left(E_m+i\frac{\gamma}{2}\right)^2}}.
\end{equation}
Multiplying Eq.~\eqref{master_station} by $\langle m|$ from the left and by $|m'\rangle$ from the right we find
\begin{equation}\label{master_station2}
 i(E_m-E_{m'}){\rho}_{m,m'}=
\frac{\epsilon^2}{4}\left(
Q_{m,m'}-\sum_{\ell=1,L}\sum_{\bar{m},\bar{m}'=1}^{L}{\mathbb R}_{m,m',\bar{m},\bar{m}'}^{(\ell)}~{\rho}_{\bar{m},\bar{m}'}\right),
\end{equation}
where the source term $Q_{m,m'}$ and the Redfield relaxation tensor ${\mathbb R}_{m,m',\bar{m},\bar{m}'}^{(\ell)}$
are given by
\begin{align}\label{Redfield}
& Q_{m,m'}=
\frac{J_{\rm r} \langle m|1\rangle\langle 1|m'\rangle}{\pi\sqrt{J_{\rm r}^2-\mu^2}}
\left(\frac{i}{E_{m'}-\mu+i\frac{\gamma}{2}}-\frac{i}{E_{m}-\mu-i\frac{\gamma}{2}}\right) ,\nonumber \\
& {\mathbb R}_{m,m',\bar{m},\bar{m}'}^{(\ell)}=\frac{\langle m|\ell\rangle\langle \bar{m}|\ell\rangle\delta_{\bar{m}',m'}}
{\sqrt{J_r^2-\left(E_{\bar{m}'}+i\frac{\gamma}{2} \right)^2}}+
\frac{\langle m'|\ell\rangle\langle \bar{m}'|\ell\rangle\delta_{\bar{m},m}}
{\sqrt{J_r^2-\left(E_{\bar{m}}-i\frac{\gamma}{2} \right)^2}}.
\end{align}

\section{Numerical Validation}
Let us assume that the left reservoir is maintained at chemical potential
$\mu+\Delta\mu$, while the right at $\mu$.
The other parameters of the reservoirs are the same if not stated otherwise.
We are interested in the stationary current across
the chain as the function of the chemical potential $\mu$ and the relaxation constant $\gamma$.
We can calculate the current by using the following approaches:

(i) by straightforward  numerical simulation
of the system's dynamics according to Eqs.~(\ref{one}-\ref{three}), which does not involve any approximation
but is very time consuming;

(ii) by simulating the system dynamics on the basis of non-Markovian master
equation, Eq.~(\ref{master_reduced}) and Eq.~(\ref{Ldag2}), which implies validity of the Born approximation, and $M\rightarrow\infty$;

(iii) by using the stationary Redfield equation Eq.~\eqref{master_station2}, and Eq.~\eqref{Redfield},
which also assumes low temperatures and the limit
$\Delta\mu\rightarrow0$; and

(iv) by applying the analytic solution Eq.~\eqref{current} which, however, implies validity of the Markov approximation.
\begin{figure*}[ht]
\centering
{\includegraphics[width=1\textwidth,height=0.56\textwidth,trim={.1cm
9.cm 0.0cm 8.1cm},clip]{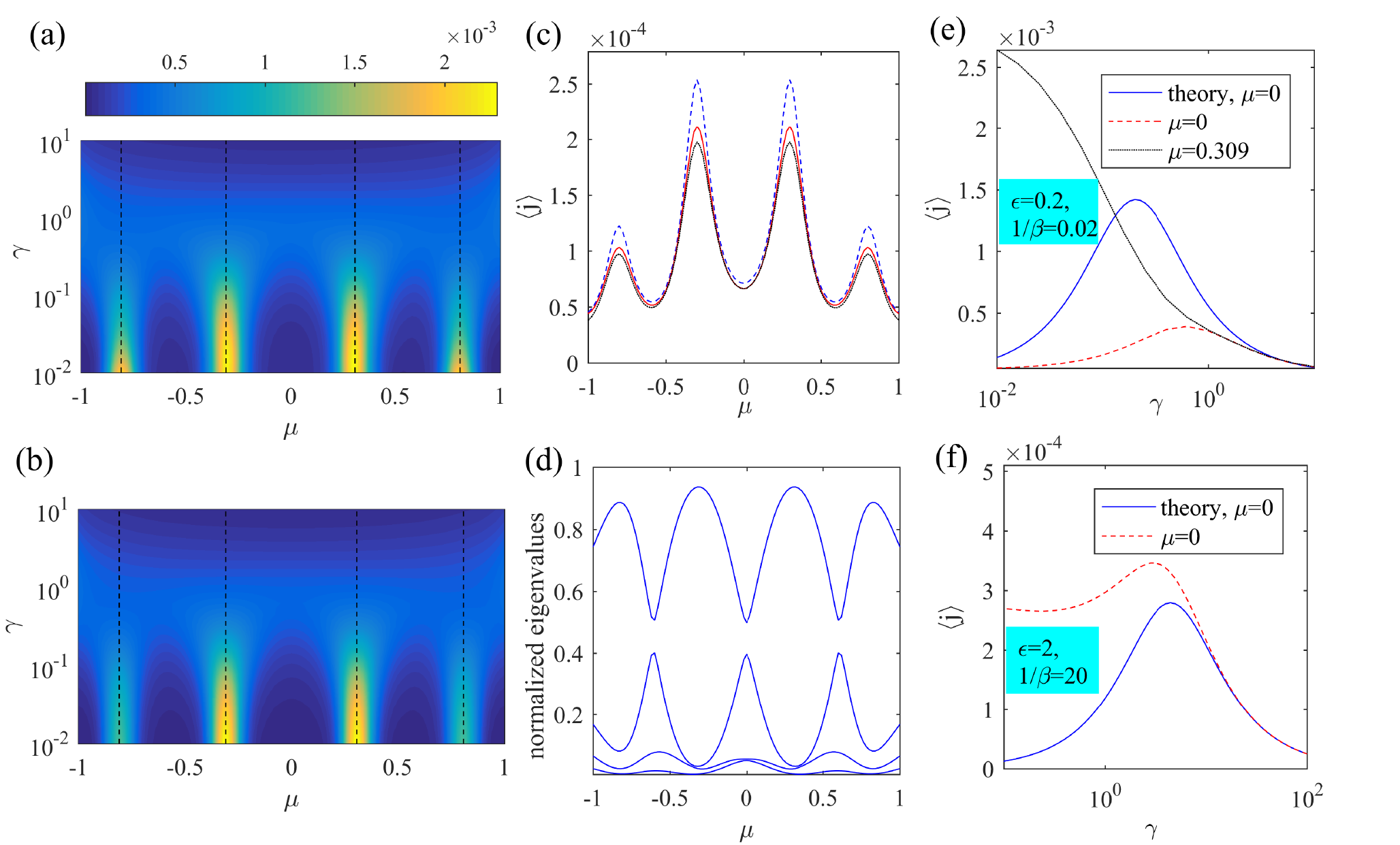}} \caption{Quantum transport of Fermionic carriers, $L=4$, $M=80$, $J_{\rm s}=1$, $J_{\rm r}=1.1$.
(a) The probability current
versus $\gamma$ and $\mu$ computed from numerical solution of Eqs.~(\ref{one}-\ref{three}), $\Delta\mu=0.1$, $1/\beta=0.02$, $\epsilon=0.2$.
The vertical dash lines show the eigenvalues of $\widehat{H}_{\rm s}$.
(b) Same
as (a) but computed by numerically solving the non-Markovian master equation, Eq.~(\ref{master_reduced}) and Eq.~(\ref{Ldag2}).
(c) The probability current against chemical potential with $1/\beta=0.02$, $\epsilon=0.2$, $\gamma=0.2$, $\Delta\mu=0.025$; dash-dot black line
shows the data obtained from numerical solution of Eqs.~(\ref{one}-\ref{three}), solid red line - from the non-Markovian master equation,
Eq.~(\ref{master_reduced}) and Eq.~(\ref{Ldag2}), and dash blue line from the stationary Redfield equation Eq.~\eqref{master_station2}, and
Eq.~\eqref{Redfield}. (d) Trace normalized eigenvalues of $\hat{\rho}_{\rm s}^{\scs (1)}$ for the same parameters as in (c).
(e-f) The probability current against $\gamma$ for $\Delta\mu=0.1$; solid lines show theoretical result, Eq.~(\ref{current}),
dashed and dotted
lines - numerical solutions.
} \label{fig2}
\end{figure*}

In Fig.~\ref{fig2}~(a, b) we compare the probability currents obtained by using (i) and (ii).
In Fig. \ref{fig2}~(a, b)  one can see a good coincidence between the two   approaches for $\mu$ spanning the whole propagation band
of $\widehat{H}_{\rm s}$ in a broad range of $\gamma$. Importantly, at small $\gamma$ we observed four resonant peaks coinciding
with the positions of the eigenvalues of $\widehat{H}_{\rm s}$, which can be explained by the onset of resonant transport due
to the coupling suppressed with small $\epsilon$. This resonant picture resembles Landauer's conductivity in which the transport solution
is the pure scattering state with the energy equal to the chemical potential \cite{Datt95}. The key to onset of the resonant
transport is $\Delta\mu$ smaller than the spacing between the eigenstates of $\widehat{H}_{\rm s}$. On the other hand a
small $\Delta\mu$ is difficult to handle with Eqs.~(\ref{one}-\ref{three}) since $\Delta\mu$ must be much larger than the level spacing
in the reservoirs which overwise would exhibit discrete eigenenergies rather than the continuous density of states Eq.~(\ref{DOS}).
In Fig. \ref{fig2} (c) we, however, managed to achieve a reasonable coincidence between (i), (ii), and (iii), where the
latter explicitly assumes infinite size reservoirs.
In Fig.~\ref{fig2}~(d) we plotted the trace-normalized eigenvalues of the transport state
$\hat{\rho}_{\rm s}^{\scs (1)}$. One can see that despite the superficial resemblance to Landauer's
conductance $\hat{\rho}_{\rm s}^{\scs (1)}$ only approaches a pure state near the resonant eigenvalues. Finally, in Figs. \ref{fig2}~(e,~f) we
compared the numerical
data with the Markovian analytic solution (iv). As expected Eq.~(\ref{current}) is only valid at large $\gamma$.
One can see from Fig.~\ref{fig2}~(e,~f) that the Markovian solution is incapable of describing the resonant transport at small $\gamma$.
Notice that Eq. (\ref{current}) is only plotted for $\mu=0$ since for $\mu=0.309$ the plots are almost identical.
As it is seen from Fig.~\ref{fig2}~(f), it is possible to approach the maximum of Eq.~(\ref{current}) by increasing $\epsilon$ and $1/\beta$, but
the Markov approximation unavoidably breaks down at $\gamma\approx J_{\rm r, s}$.

\begin{figure*}[t]
\centering
{\includegraphics[width=0.95\textwidth,height=0.34\textwidth,trim={.1cm
11.cm 0.0cm 11.1cm},clip]{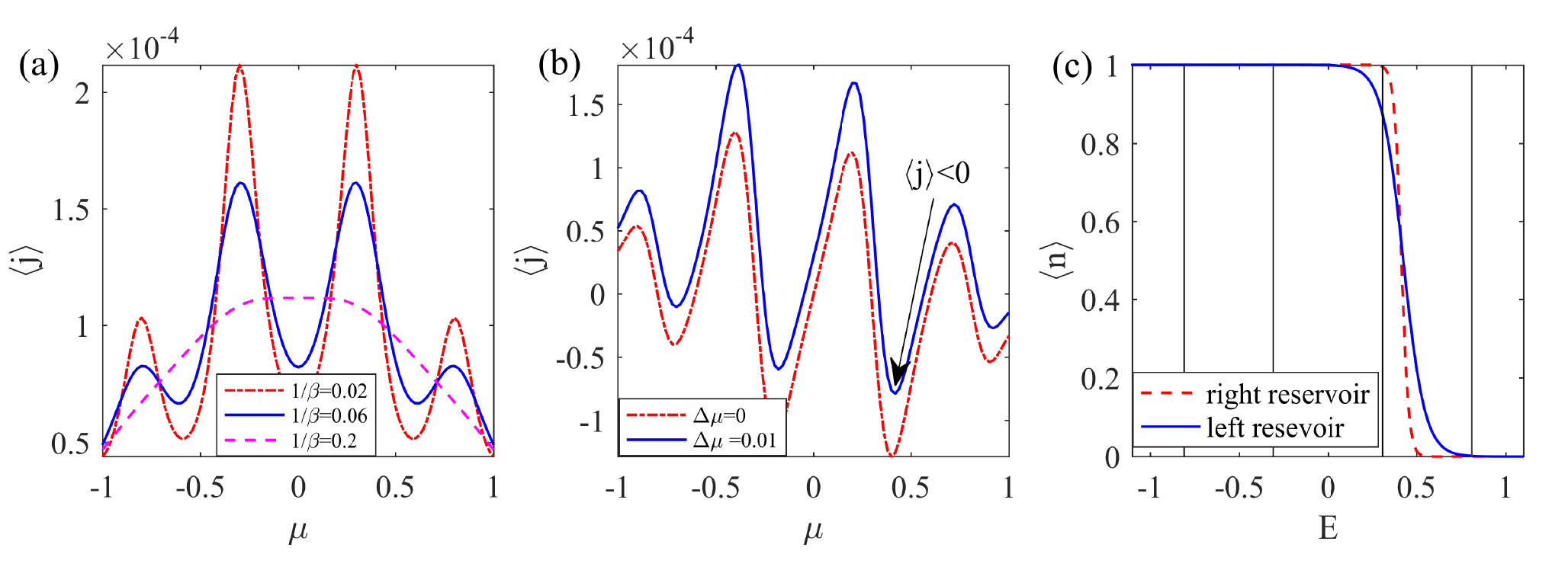}} \caption{Temperature effect on transport, $L=4$, $M=80$, $J_{\rm s}=1$, $J_{\rm r}=1.1$, $\gamma=0.2$,
$\epsilon=0.2$.
(a) The effect of increasing temperature in both reservoirs, $\Delta\mu=0.025$.
(b) The effect of different temperatures; temperature in the left reservoir $1/\beta=0.06$, temperature in
the right reservoir $1/\beta=0.02$. (c) Fermi-Dirac distributions in the reservoirs for subplot (b) with $\mu=0.42$ and $\Delta\mu=0.01$. Vertical black
lines show the eigenvalues of $\widehat{H}_{\rm s}$.}
\label{fig3}
\end{figure*}

Let us now examine the effect of temperature on quantum transport in more detail. In Fig.~\ref{fig3}~(a) we show the dependance of
the current on the chemical potentials for three different temperatures for a fixed $\Delta\mu$. One can see in Fig.~\ref{fig3}~(a)
that the temperature increase eliminates the resonant transport. This effect could be easily understood by smoothing of the Fermi
distribution at larger temperature so that a significant difference between the Fermi functions of the reservoirs
occurs at a broader range of energies. In Fig.~\ref{fig3}~(b) we present the data for the case of temperature increase
only in the left reservoir. Notice that even with a larger chemical potential of the left reservoir
$\Delta\mu=0.01$ one can observe a flow of particles from the right to the left $\langle{j}\rangle<0$ at $\mu=0.42$.
To understand this effect
in Fig.~\ref{fig3}~(c) we plotted the Fermi-Dirac distributions in both reservoirs superposed
with the eigenvalues of $\widehat{H}_{\rm s}$. One can see that all but the third eigenvalue
occur at the points where the populations are almost equal. The third eigenvalue, though, occurs in
the point where the population in the right reservoir is larger leading to enhancement of the
resonant transport from the right to the left.

\section{Summary}
In summary, we analyzed a fermionic model that allows for elementary derivation of transport master equation which can be solved analytically
in the Markov approximation.
In the Born approximation we have derived a non-Markovian master equation for quantum transport of
fermionic carriers in the Redfield form. The equation obtained is shown to predict the effect of resonant transport which can not be
accounted for by the exact Markovian solution.
A similar phenomenon of resonant excitation transfer has been recently predicted in \cite{Chetcuti20} in the
set-up consisting of a block of sender sites,
each hosting one excitation, weakly coupled to a quantum tight-binding wire at one edge with the block of receiver sites weakly
coupled at the opposite edge. Here we have demonstrated the effect of resonant transport with thermalized sender and receiver.
Two ingredients are essential for the correct derivation
of the non-Markovian master equation. First, the explicit account of the relaxation rate in the microscopic model of the
reservoir. The reservoir relaxation rate is found to be of key importance to ensure convergence of the memory integral and
onset of the resonant transport.
The second ingredient is the account of correlations between the state of the reservoir and the state of the system. Such correlations
do not allow to write the total density matrix as the tensor product of the density matrices of the reservoir and the system,
but, nonetheless, do not invalidate the Born approximation at weak couplings. We speculate that the above conclusion can be applied
to quite arbitrary set-ups of reservoir-coupled conductors. The benefit of the set-up considered, though, is the simplicity of
derivation that paves a way to generalizing the results for interparticle interactions
in the system.

We acknowledge financial support from Russian Science Foundation through Grant No. 19-12-00167.

\section*{References}
\bibliographystyle{unsrt}
\bibliography{Open_systems}

\appendix
\section{} 
\numberwithin{equation}{section}
In Eq. (21) and Eq. (22) of the main text we arrived at
the following equation for the system's SPDM
\begin{align}\label{S2}
& \frac{\partial \hat{\rho}_{\rm s}}{\partial t}=-i[\widehat{H}_{\rm s},\hat{\rho}_{\rm s}]
+\sum_{\ell=1,L}\left(\widehat{L}_{\ell}+
\widehat{L}_{\ell}^{\dagger}\right), \nonumber \\
& \widehat{L}_{\ell}\!=\!\epsilon^2\!\int\limits_{-t}^{0}d\tau e^{\frac{\gamma}{2}\!\tau}
\widehat{V}^{\dagger}_{\ell}\widehat{U}_{\rm r}^{\dagger}(t)\left[
\hat{\rho}_{\ell}(\tau\!+\!t)\widehat{V}_{\ell}\!-\!\widehat{V}_{\ell}\hat{\rho}_{\rm s}(\tau\!+\!t)\right]\widehat{U}_{\rm s}(t).
\end{align}
In the similar fashion Eq.~(18) and Eq.~(19) of the main text can be resolved to
\begin{align}\label{S3}
& \frac{\partial \hat{\rho}_{{\ell}}}{\partial t}=-i[\widehat{H}_{\rm r},
\hat{\rho}_{\ell}]
+\widehat{K}_{\ell}+
\widehat{K}_{\ell}^{\dagger} +\gamma(\hat{\rho}^{\scs (0)}_{\ell}-\hat{\rho}_{\ell}), \nonumber\\
& \widehat{K}_{\ell}\!=\!\epsilon^2\!\int\limits_{-t}^{0}d\tau
e^{\frac{\gamma}{2}\!\tau}
\widehat{U}_{\rm r}^{\dagger}(t)\left[\widehat{V}_{\ell}
\hat{\rho}_{\rm s}(\tau\!+\!t)\!-\!
\hat{\rho}_{\ell}(\tau\!+\!t)\widehat{V}_{\ell}\right]\widehat{U}_{\rm s}(t)
\widehat{V}^{\dagger}_{\ell}, \nonumber \\
& \hat{\rho}^{\scs (0)}_{ \rm \ell}=\sum_{\nu=1}^M\frac{|\nu\rangle\langle\nu|}{e^{-\beta[J_{\rm r}\cos(k_{\nu})+\mu]}+1},
\end{align}
where subscript $\ell$ specifies the reservoir's connection site.
Together Eq.~(\ref{S2}) and Eq.~(\ref{S3}) constitute a set of tree integro-differential equations for
$\hat{\rho}_{\rm s}$ and $\hat{\rho}_{\rm \ell}$ with $\ell=1,L$.

 The Markov approximation consists of assuming no memory in integrals in Eq.~(\ref{S2}) and Eq.~(\ref{S3}).
 It can be applied under two assumptions:
\begin{itemize}
\item $\gamma\gg J_{\rm r}, J_{\rm s}$; i.e. the reservoirs' relaxation rate is much greater
than the characteristic dynamic time-scales due to the evolution operators
$\widehat{U}_{\rm s, r}(t)$ of both reservoirs and system.
\item The reservoirs and the system are near stationary. Thus, $\hat{\rho}_{\rm s}$ and $\hat{\rho}_{\rm \ell}$
are slow varying on the scale $1/\gamma$.
\end{itemize}
The memory effect can be
removed by applying
\begin{equation}
\int\limits_{-t}^{0}d\tau e^{\frac{\gamma}{2}\tau}\widehat{A}(\tau+t)=\frac{2}{\gamma}\widehat{A}(t)
\end{equation}
where $\widehat{A}(t)$ is any operator quantity slow varying on the scale $1/\gamma$.
Under the Markov approximation one finds
\begin{align}\label{LMdagM}
& \widehat{L}^{\dagger}_{\ell}=2\frac{\epsilon^2}{\gamma}
\widehat{V}^{\dagger}_{\ell}\left(
\hat{\rho}_{\ell}\widehat{V}_{\ell}-\widehat{V}_{\ell}\hat{\rho}_{\rm s}\right), \nonumber \\
& \widehat{K}^{\dagger}_{\ell}=2\frac{\epsilon^2}{\gamma}
\left(\widehat{V}_{\ell}\hat{\rho}_{\rm s}-\hat{\rho}_{\ell}\widehat{V}_{\ell}\right)
\widehat{V}^{\dagger}_{\ell}.
\end{align}
Substituting the above into the first line Eq.~(\ref{S2}) one finds
\begin{align}\label{master_res2}
\frac{\partial \hat{\rho}_{\ell}}{\partial t}\!=\!-i[\widehat{H}_{\rm r},
\hat{\rho}_{\ell}]\!+\!\gamma(\hat{\rho}^{\scs (0)}_{\ell}\!-\!\hat{\rho}_{\ell})\!-
\!\frac{2\epsilon^2}{\gamma}\!\left(\!\{\widehat{V}_{\ell}\widehat{V}_{\ell}^{\dagger}\!,\hat{\rho}_{\ell}\!\}
\!-\!2\widehat{V}_{\ell}\hat{\rho}_{\rm s}\widehat{V}_{\ell}^{\dagger}\!\right)\!,
\end{align}
where $\{{\scs \ldots},{\scs \ldots}\}$ designates the anticommutator.
With the initial condition $ \hat{\rho}_{\ell}(0)=0$ and $J_{\rm r}\ll \gamma$
the solution of Eq. (\ref{master_res2}) reads
\begin{equation}\label{S5}
\hat{\rho}_{\ell}=\int\limits_{-t}^{0}d\tau \widehat{F}(\tau)
\left[\gamma\hat{\rho}^{\scs (0)}_{ \ell}+
4\frac{\epsilon^2}{\gamma}\widehat{V}_{\ell}\hat{\rho}_{\rm s}(\tau+t)\widehat{V}_{\ell}^{\dagger}\right]
\widehat{F}(\tau),
\end{equation}
where
\begin{equation}
\widehat{F}(t)=e^{
2\frac{\epsilon^2}{\gamma}\widehat{V}_{\ell}\widehat{V}_{\ell}^{\dagger}
t}.
\end{equation}
Notice that the initial condition $ \hat{\rho}_{\ell}(0)=0$ is far from the thermodynamic equilibrium
which seemingly contradicts our initial assumptions. Yet, for large times $t\gg1/\gamma$, when the equilibrium
in the isolated reservoirs is settled, the initial condition for the operator $\widehat{K}_{\ell}$ in Eq.~(\ref{S3}) becomes irrelevant  with
all deviations from the equilibrium due to the coupling with the system that is accounted for exactly in
both Eq.~(\ref{master_res2}) and Eq.~(\ref{S5}).
By using the definition of the coupling operator
\begin{equation}\label{Vap}
\widehat{V}_{\ell}=-\frac{1}{2\sqrt{M}}\sum_{\nu=1}^M|\nu\rangle\langle\ell|.
\end{equation}
one finds
\begin{align}
\widehat{F}(t)
=\widehat{\mathbb I}_{\rm r}+\frac{e^{\frac{\epsilon^2}{2\gamma}t}-1}{M}\sum^M_{\nu, \nu'=1}|\nu\rangle\langle \nu'|.
\end{align}
Now the quantity $\widehat{V}^{\dagger}_{\ell}
\hat{\rho}_{\ell}\widehat{V}_{\ell}$ that has emerged in Eq. (\ref{LMdagM}) can be written as
\begin{equation}\label{step1}
\widehat{V}^{\dagger}_{\ell}
\hat{\rho}_{\ell}\widehat{V}_{\ell}=\!\int\limits_{-t}^{0}d\tau\frac{e^{\frac{\gamma^2+2\epsilon^2}{\gamma}\tau}}{4}
\!\left[\,\gamma\bar{n}_{\ell}\!+\!\frac{\epsilon^2}{\gamma}\langle \ell|\hat{\rho}_{\rm s}(\tau\!+\!t)|\ell \rangle
\!\right]\!|\ell \rangle\langle \ell|,
\end{equation}
where $\bar{n}_{\ell}$ is the mean population of each site of the reservoir at the $\ell_{\rm th}$ site
in the absence of coupling $\epsilon\!=\!0$
\begin{equation}
\bar{n}_{\ell}=\frac{1}{M}\sum_{\nu=1}^M\frac{1}{e^{-\beta[J_{\rm r}\cos(k_{\nu})+\mu]}+1}.
\end{equation}
Assuming no memory again one rewrites Eq. (\ref{step1}) as
\begin{equation}\label{step2}
\widehat{V}^{\dagger}_{\ell}\hat{\rho}_{\ell}\widehat{V}_{\ell}=\frac{1}{4}
\left(\frac{{\gamma^2\bar{n}_{\ell}}}{\gamma^2+\epsilon^2}+\frac{{\epsilon^2}}{\gamma^2+
\epsilon^2} \langle \ell|\hat{\rho}_{\rm s}|\ell \rangle \right)|\ell \rangle\langle \ell|.
\end{equation}
Finally, by combining Eqs. (\ref{S2}, \ref{LMdagM}, \ref{step2}) we arrive at the
Markovian master equation for the system's SPDM
\begin{align}\label{Markovianap}
& \frac{\partial \hat{\rho}_{\rm s}}{\partial t}=-i[\widehat{H}_{\rm s}, \hat{\rho}_{\rm s}]
-\frac{\epsilon^2}{2\gamma}\sum_{\ell=1,L}\left\{|\ell \rangle\langle \ell|,\hat{\rho}_{\rm s} \right\}
\nonumber \\
&
+\frac{\epsilon^2}{\gamma}\sum_{\ell=1,L}\left(\frac{{\gamma^2\bar{n}_{\ell}}}{\gamma^2+\epsilon^2}+
\frac{{\epsilon^2}}{\gamma^2+\epsilon^2} \langle \ell|\hat{\rho}_{\rm s}|\ell \rangle \right)
|\ell \rangle\langle \ell|.
\end{align}

\end{document}